 \documentclass[aps,amsmath,amssymbd,dvips,preprint,showkeys,preprint]{revtex4}

\usepackage{graphicx}
\usepackage{graphics}
\usepackage{color}
\def \be  {\begin{equation}}
\def \ee  {\end{equation}}
\def \ee  {\end{equation}}
\def \bea {\begin{eqnarray}}
\def \eea {\end{eqnarray}}

\def \Tr  {\bf{Tr}}

\def \ra  {\rightarrow}

\begin{document}

\preprint{ECTP-2013-18\hspace*{0.5cm}and\hspace*{0.5cm}WLCAPP-2013-15}

\title{Thermal Description of Particle Production in Au-Au Collisions at STAR Energies}
\author{Abdel~Nasser~TAWFIK\footnote{http://www.atawfik.net/}}
\email{a.tawfik@eng.mti.edu.eg}
\affiliation{Egyptian Center for Theoretical Physics (ECTP), MTI University, 11571 Cairo, Egypt}
\affiliation{World Laboratory for Cosmology And Particle Physics (WLCAPP), Cairo, Egypt}

\author{Ehab~ABBAS}
%\email{e.gamal@eng.mti.edu.eg}
\affiliation{Egyptian Center for Theoretical Physics (ECTP), MTI University, 11571 Cairo, Egypt}
\affiliation{World Laboratory for Cosmology And Particle Physics (WLCAPP), Cairo, Egypt}

\date{\today}

\begin{abstract}

The hadron ratios measured in central Au-Au collisions are analysed by means of Hadron Resonance Gas (HRG) model over a wide range of nucleon-nucleon center-of-mass energies, \hbox{$\sqrt{s_{NN}}=7.7-200~$GeV} as offered by the STAR Beam Energy Scan I (BES-I). We restrict the discussion on STAR BES-I, because of large statistics and over all homogeneity of STAR measurements (one detector) against previous experiments. Over the last three decades, various heavy-ion experiments utilizing different detectors (different accuracies) have been carried out. Regularities in produced particles at different energies haven been studied. The temperature and baryon chemical potential are deduced from fits of experimental ratios to thermal model calculations assuming chemical equilibrium. We find that the resulting freeze-out parameters using single hard-core value and point-like constituents of HRG are identical. This implies that the excluded-volume comes up with no effect on the extracted parameters. We compare the results with other studies and with the lattice QCD calculations. Various freeze-out conditions are confronted with the resulting data set. The effect of feed-down contribution from week decay and of including new resonances are also analysed. At vanishing chemical potential, a limiting temperature was estimated, \hbox{$T_{lim}=158.5\pm 3~$MeV}. 
\end{abstract}. 


\maketitle

%%%%%%%%%%%%%%%%%%%%%%%%%%%%%%%%%%%%%%%%%%%%%%%%%%%%%%%%%%%%%%%%%%%%%%
%%%   Section I
%%%%%%%%%%%%%%%%%%%%%%%%%%%%%%%%%%%%%%%%%%%%%%%%%%%%%%%%%%%%%%%%%%%%%%

\section{Introduction}
    
    The heavy-ion experiments are carried out to study hadronic matter under extreme conditions  of high temperature or density (or both)~\cite{Cabibbo}. These conditions likely exist in the core of compact stellar objects and should be established a few microseconds after the Big Bang, when matter was in its primordial state; a soup of Quark-Gluon Plasma (QGP). Data from the Relativistic Heavy-Ion Collider (RHIC) has shown that QGP, the new state of matter, was created in Au-Au collisions~\cite{Gyulassy}. It is conjectured that the created hot and dense partonic matter rapidly expands and cools down. On path of this evolution, it undergoes phase transition(s) back to the hadronic matter. Different {\it equilibrium} thermal models~\cite{Andronic,Becattini,Cleymans,Tawfik:2005gk,Tawfik:2010pt,Torrieri,Tiwari,EVC} can very well reproduce the particle abundances, which are governed in final state - at chemical equilibrium - by two parameters, the chemical freeze-out temperature $T_{ch}$ and the baryon chemical potential $\mu_b$, where the latter reflects the net baryon content of the system. Different values of $T_{ch}$ and $\mu_b$ at different energies set up the chemical freeze-out line. This line appears very close to the phase boundary between QGP and hadronic phase, especially at low $\mu_b$. The separation between the two boundaries increases with increasing $\mu_b$.   
      
    To explore the freeze-out diagram, preciously, gaps in the energies between old RHIC at BNL  and top Superproton Synchrotron (SPS) at CERN should be closed. Therefore, the Solenoidal Tracker At RHIC (STAR) launched first phase of the Beam Energy Scan (BES-I) program~\cite{STAR1}, in order to collect data from $Au+Au$ collisions at center-of-mass energies of $39$, $27$, $19.6$, $11.5$ and $7.7~$GeV covering a wide range of baryon chemical potential $\sim100-400~$MeV in the Quantum Chromodynamics (QCD) phase diagram \cite{STAR1}. Within this energy range, two important phenomena are conjectured to be populated, the critical endpoint and closeness between chemical freeze-out boundary and QCD phase transition form QGP to hadrons (or back)~\cite{Karsch}. 

    In the present work, the freeze-out parameters, $T_{ch}$ and $\mu_b$, are extracted from fits of the experimental particle ratios with corresponding ratios calculated in the HRG model assuming chemical equilibrium. The experimental hadron ratios are limited to mid-rapidity central Au-Au collisions at energies $200$, $130$, $62.4$, $39$, $11.5$, $7.7~$GeV. At BES-I missing energies, to authors' best knowledge, there are no published results, yet. 
    
     The contributions of weak decay feed-down and effects of cut-off on hadron resonances have been studied. The results are compared with recent lattice QCD calculations, in which a limiting temperature has been estimated at vanishing chemical potential, $T_{lim}=154 \pm 9 MeV $. The lattice calculations have been performed for $2+1$ quark flavors with physical masses in the continuum limit of the chiral susceptibility i.e., the derivative of chiral condensate with respect to light quark masses when $mq \ra 0$~\cite{LQCD}.
    
   The present paper is organized as follows. Section \ref{sec:hrg} elaborates details about the HRG model . The hadron Interactions, especially excluded-volume correction shall be discussed in section \ref{sec:exclV}. A list of argumentation why we concentrate the analysis to STAR BES-I shall be outlined in section \ref{sec:ySTAR}. The fits of the experimental ratios with the HRG calculations are discussed in section \ref{sec:phys}. Section \ref{sec:res} is devoted to the results and discussion. The conclusions and outlook shall be summarized in section \ref{cons}.

\section{The hadron resonance gas model}
\label{sec:hrg}

   According to Hagedorn, the hadronic phase is assumed to have large degrees of freedom~\cite{Hagedorn}. Additionally, there should be a kind of equilibrium. Therefore, statistical mechanics can be used  in describing this phase. In nuclear collisions, the formation of  hadrons at chemical freeze-out temperature $T_{ch}$ should be controlled by the phase space and conservation laws. The phase space of each particle depends on the mass, energy, degeneracy and available volume. For large number of produced particles, grand canonical ensemble (GCE) is justified. Thus, it is straightforward to derive an expression for the pressure. In light of this, the hadron resonances can be treated as a free gas~\cite{Karsch:2003vd,Karsch:2003zq,Redlich:2004gp,Tawfik:2004sw,Tawfik:2004vv}, as the resonances are conjectured to add to the thermodynamic pressure in the hadronic phase (below $T_c$). This statement is likely valid for free as well as for strongly interacting resonances. It has been shown that the thermodynamics of strongly interacting  system can also be approximated to an ideal gas composed of hadron resonances ~\cite{Tawfik:2004sw,Vunog}. The grand canonical partition function reads
\bea
Z(T, \mu, V) &=&\Tr\left[ \exp^{\frac{\mu\, N-H}{T}}\right],
\eea
where $H$ is the Hamiltonian of the system and $T$ ($\mu$) being temperature (chemical potential). The Hamiltonian is given by the sum of the kinetic energies of relativistic Fermi and Bose particles. The main motivation of using this Hamiltonian is that it contains all relevant degrees of freedom of confined and strongly interacting matter. It includes implicitly the interactions that result in resonance formation. In addition, it has been shown that this model can submit a quite satisfactory description of particle production in heavy-ion collisions. With the above assumptions the dynamics the partition function can be calculated exactly and be expressed as a sum over {\it single-particle partition} functions $Z_i^1$ of stable hadrons and their resonances.
\bea \label{eq:lnz1}
\ln Z(T, \mu ,V)&=&\sum_i \ln Z^1_i(T,V)=\sum_i\pm \frac{V g_i}{2\pi^2}\int_0^{\infty} p^2 dp \ln\left\{1\pm \exp\left[\frac{\mu_i -\varepsilon_i(p)}{T}\right]\right\},
\eea
where $\varepsilon_i(p)=(p^2+ m_i^2)^{1/2}$ is the $i-$th particle dispersion relation, $g_i$ is spin-isospin degeneracy factor and $\pm$ stands for bosons and fermions, respectively. The $i$-th particle chemical potential is given as \hbox{$\mu_i=\mu_b B_i+\mu_s S_i+\mu_{I^3} I_i^3$}, where $\ B_i, S_i $  and  $I_i^3$ are the baryon, strange and isospin quantum number, respectively. The thermodynamic properties of the system can be obtained from the partition function of all resonances in the hadronic phase. 

At finite temperature $T$ and baryon chemical potential $\mu_i$, the pressure of the $i$-th hadron or resonance species reads 
\begin{equation}
\label{eq:prss} 
p(T,\mu_i ) = \pm \frac{g_i}{2\pi^2}T \int_{0}^{\infty}
           p^2 dp  \ln\left\{1 \pm \exp\left[\frac{\mu_i -\varepsilon_i(p)}{T}\right]\right\}.
\end{equation} 
As no phase transition is conjectured in HRG, summing over all hadron resonances results in the final thermodynamic pressure. The number density can be obtained as
\bea \label{eq:n_i}
n(T, \mu) &=& \sum_i \frac{\partial}{\partial \mu_i} p(T,\mu_i) = \sum_i \frac{g_i}{2\, \pi^2}\, \int_0^{\infty}\frac{p^2 dp}{ \exp\left[\frac{\mu_i -\varepsilon_i(p)}{T}\right]\pm 1}.
\eea 

The conservations laws should be fulfilled through out the chemical potentials and temperature and over the complete phase space. They include conservation of strangeness \hbox{$V \sum_i\ n_i(T, \mu_i) S_i =0$} (vanishing strangeness),  baryon, \hbox{$V \sum_i\ n_i(T, \mu_i) B_i = Z+N$} (conserved charge and baryon number) and  isospin, \hbox{$V \sum_i\ n_i(T, \mu_i) I_i^3=(N-Z)/2$}, where $N$ and $Z$ are the neutron and protons number in the colliding nuclei.
   
For completeness, it is worthwhile to mention that the degree of non-equilibrium can be implemented - among others - through the strange quark occupation factor $\gamma_s$ (and may be also that of light quarks $\gamma_q $) \cite{gammaS1,gammaS2} in the partition function
\bea \label{eq:lnzallG}
\ln Z(T, \mu, V,\gamma_s) &=& \sum_i\pm \frac{V\, g_i}{2\, \pi^2}\int_0^{\infty} p^2 dp\, \ln\left(1\pm \gamma_s^{s_i}\exp\left[\frac{\mu_i -\varepsilon_i(p)}{T}\right]\right),
\eea 
where $s_i$ is number of strange valence quarks or antiquarks in the $i$-th hadron. The value of $\gamma_s$ is always less than unity~\cite{gammaS1,Andronic}. This apparently points to strangeness phase space suppression.   
  
During the final expansion, we assume that inelastic interactions between resonances and annihilation process~\cite{SHMUrQM} have negligible contributions to the final state. The main process at this stage is unstable resonance decay. So, the final number density for $i$-th particle is given as 
\bea \label{eq:n_i^{final}}
\ n_i^{final} &=& n_i + \sum_j\ Br_{j\ra i} n_j,
\eea 
where $Br_{j\ra i}$ is the effective branching ratio of $j$-th hadron resonance into $i$-th particle. Taking into consideration all multi-step decay cascades, then 
\bea \label{eq:Br}
\ Br_{j\ra i} &=& br_{j\ra i} + \sum_{l_1}\ br_{j\ra l_1}br_{l_1\ra i}+ \sum_{l_1,l_2}\ br_{j\ra l_1}br_{l_1\ra l_2} br_{l_2\ra i}+ \cdots,
\eea 
where the $br_{j\ra i}$ is the branching ratio $j$-th hadron resonance into $i$-th hadron.
       
The switching between hadron and quark chemistry is given by the relations between  the hadronic chemical potentials and the quark constituents, \hbox{$\mu_i =3\, B_i\, \mu_q + S_i\, \mu_S +\mu_{I^3} I_i^3$}. The chemical potential assigned to the strange quark read $\mu_S=\mu_q-\mu_s$ and to the light quarks is $\mu_q=(\mu_u+\mu_d)/2$. The latter is only, if iso-spin chemical potential vanishes. Otherwise, $\mu_{I^3}=(\mu_u-\mu_d)/2$. The strangeness and isospin chemical potential are calculated as a function of $T$ and $\mu_i $ under the conservation laws in heavy-ion collisions~\cite{Tawfik:2004sw}.
     
       In the present work, we include contributions of the  hadrons which consist of light and strange quark flavors listed in the most recent PDG~\cite{PDG}. This corresponds to $388$ different isospin states of mesons and baryons besides their anti-particles. The decay branching ratios are also taken from Ref.~\cite{PDG}. For the observed decay channels with unknown probabilities, we follow the rules given in Ref.~\cite{MichalecPh.D}. 
       
       Because the inclusion of finite volume at energies higher than that of the Alternating Gradient Synchrotron (AGS), is not applicable, the zero-width approximation is utilized. An additional reason for this is the large numerical costs and the less improvement in the final results~\cite{Andronic,Michalec1,Petran}. The excluded-volume correction (EVC)~\cite{EVC} is applied taking into account the volume occupied by individual hadrons with radii $r_m$ for mesons and $r_b$ for baryons. The thermodynamic quantities are conjectured to be modified due to EVC. The corrected pressure will be obtained by an iterative procedure, 
 \bea 
 p^{excl}(T, \mu_i) &= p^{id}(T, \tilde{\mu_i}), ~~~~~~~~~~~{\tilde{\mu_i}} &= \mu_i - \upsilon p^{excl}(T, \tilde{\mu_i}), \label{eq:P}
\eea 
where $p^{id} (p^{excl})$ being the pressure in the ideal case (case of excluded volume) and $\upsilon$  is the {\it eigen} volume which is calculated for a radius, $16 \pi r^3/3$~\cite{Landau}. The framework of GCE and assuming full chemical equilibrium, i.e. $\gamma_s = 1$, is the one which we utilize in the present analysis. 
      
\section{Hadron Interactions and Excluded-Volume Correction}
\label{sec:exclV}

There are various types of interactions to be implemented to the {\it ideal} hadron gas. Besides van der Waals repulsive interaction, Uhlenbeck and Gropper statistical correction and strong interactions (s-matrix) should be estimated. For a recent review, readers can consult Ref. \cite{Tawfik:2013eua}. In the present work, we concentrate the discussion on the van der Waals repulsive interaction. The repulsive interactions between hadrons are considered as a phenomenological extension, which would be exclusively based on van der Waals excluded volume \cite{EVC}. Assuming that hadrons have spherical hard-core, a considerable modification in the thermodynamic quantities of the ideal hadron gas likely takes place. The main problem in this correction is the hard-core, where a clear experimental evidence would be originated in from nucleon-nucleon scattering. Experiments confirmed that the proton should have a hare-core of radius of $0.3~$fm. The key question is related to the other resonances? In literature \cite{Tiwari,EVCR}, we find a wide spectrum of hard-core radii ranging from $ 0.0$ and $0.8~$fm.     

As an attempt to solve this problem, Tawfik \cite{Tawfik:2013eua} confronted various thermodynamics quantities calculated in HRG with different hadron radii to the first principle lattice QCD simulations \cite{latFodor}. The latter offers an essential framework to check the ability of extended {\it non-ideal} hadron gas, in which the excluded volume is taken into consideration \cite{apj}, to describe the hadronic matter in thermal and dense medium. It has been concluded that increasing the hard-core radius reduces the ability to reproduce the lattice QCD calculations. At $0\leq r<0.2~$fm, the ability of HRG model to reproduce the lattice energy density or trace anomaly is very high. The three radii, $r=[0.0,0.1,0.2]~$fm have almost the same results. At $r>0.2~$fm, the disagreement  becomes obvious and increases with increasing $r$. Thus, we restrict the hard-core radius to small values. 

In the present work, we estimate the freeze-out parameters in HRG with point-like and finite hard-core constituents, section \ref{sec:phys}. The results at $r=0$ is almost identical to the ones at $r=0.3~$fm, Tab. \ref{tab:1}. With $r=0.3~$fm, we mean that both mesons and bosons have the same radius. This implies that {\it the excluded volume is practically irrelevant when extracting the chemical freeze-out parameters}. As discussed in next sections, different radii assigned to mesons and baryons lead to less convincing results.
   
\section{Why STAR experiment?}
\label{sec:ySTAR}

In this section, we list out argumentation why we concentrate the analysis to STAR BES-I. STAR is one of the two large detector systems constructed at RHIC at the Brookhaven National Laboratory (BNL). It was designed to investigate strongly interacting matter and search for signatures of Quark-Gluon Plasma (QGP) and QCD phase transition. STAR consists of several types of detectors, each specializing in detecting certain types of particles. These detectors should work together in an advanced data acquisition and subsequent physics analysis that allows final statements to be made about the collision. In many physics experiments, a theoretical idea can be tested directly by a single measurement. STAR was the first in making use of a variety of simultaneous studies in order to draw strong conclusions about the underlying physics. The complexity of the system formed in the high-energy nuclear collision and the unexplored landscape of the physics to be studied are the reasons. 
\begin{itemize}
\item First, the charged hadron identification at high $p_T$ can be done with STAR TPC (TOF). Particle IDentification (PID) capabilities use Time Projection Chamber (TPC) and a Time-Of-Flight (TOF) detector. The identification capability of charged hadrons is greatly extended compared with that achieved by TPC and TOF, separately. 
\item Second, the data from Phase-I of the BES program offers a unique opportunity to scan the QCD phase diagram and even map out the chemical freeze-out boundary using homogeneous data set. 
\item Third, the problem of taking unclear specified weak decay contribution in PHENIX~\cite{PHENIX} can be avoided. Otherwise, we should assume different types of contributions~\cite{Andronic}. BES-I would replace top SPS energies. Irregularities registered in SPS measurements led to large differences in the extracted freeze-out parameters. This can been noticed, for instance, using NA49, NA44, and NA57 measurements at $17.3~$GeV ~\cite{Andronic}. 
\end{itemize}

Firstly, the STAR BES-I enables measurements to be made at energies ranging from SPS to top RHIC with the same detector. Secondly, the same uniform acceptance likely occurs at each energy value. Finally, not only the statistics gets better due to the higher acceptance of the STAR detector, but also, there will be cleaner and more extensive PID capabilities. This allows not only to repeat the SPS measurements in much finer details but also to enhance them through differential and multiple measures. The latter is very essential, since integrating measures because of small statistics or limited acceptance is connected with lost in valuable information.

\section{Statistical Fits with STAR particle ratios}
\label{sec:phys}

    From $N$ experimental hadron yields, $N-1$ statistically independent ratios can be constructed. Assuming full chemical equilibrium in thermal models, just two parameters should be estimated in order to reproduce the experimental data. The present work is devoted to these thermal parameters, $T_{ch}$ and $\mu_b$. Therefore, fitting the HRG results with the statistically-independent experimental ratios determines the statistically-best parameters. Other methods such as the fitting to the measured hadron yields~\cite{ yields}, can be used to deduce another parameter, the fireball volume. This might be included in a future work. Unstable resonance states like $\phi$ and $ k^{*}$ will be avoided, because of re-scattering and regeneration~\cite{unstableR}, which likely happened during the later expansion after the chemical freeze-out.
      
    At RHIC energies, the rapidity distribution exhibits a boost-invariant plateau near mid-rapidity~\cite{Ullrich}. As demonstrated in Ref.~\cite{CleymansRapi}, the effects of hydrodynamic flow would be cancelled out in the particle ratios. Therefore, the measurements at mid-rapidity are consistent with the framework of the HRG model, which does not contain any dynamical treatment. For our analysis, the most-central collisions are strongly recommended, especially when the statistical treatment is based on GCE. The contributions of weak decays should be implemented in the HRG model, in order to match with the experimental conditions.
    
    The analysis includes $11$ (occasionally $10$) independent particle ratios. Additional particle ratios can also be included, as soon as these are available. The number of particle ratios is kept unchanged at all energies. The best fits are assured to take place in the $11$-independent-particle-ratios, simultaneously. In order words, we did not fit individual particle ratios. Furthermore, this rule is respected at all energies. Another rule governing the present analysis is that the hadron ratios - with corresponding errors - are taken from the STAR experiment. In some cases, the ratios - and their corresponding errors - have been calculated from the published yields, wherever they are available. Calculating particle ratios using STAR yields was enforced, as no experimentally-estimated particle ratios were available.   
    
The criterion for the best statistical fitting is based on estimating minima, for instance
\bea \label{eq:chi}
\chi^2 &=& \sum_i\frac{\left(R_i^{exp}-R_i^{model}\right)^2}{\sigma^2},
\eea 
and quadratic deviation 
 \bea \label{eq:q}
  q^2 &=& \sum_i\frac{\left(R_i^{exp}-R_i^{model}\right)^2}{\left(R_i^{model}\right)^2},
\eea 
where $R_i^{exp}$ ($R_i^{model}$) is the $i$-th measured (calculated) ratio and $\sigma$ is the experimental data errors. Both methods, $\chi^2$ and $q^2$, are implement. Any possible deviation between these two methods should give an indication about the accuracy of the parameters extracted from the given data set. The statistically-independent ratios used in estimating $\chi^2$ and $q^2$ at $200$, $130$ and $62.4$ GeV are $\pi^-$/$\pi^+$, $k^-$/$k^+$, $\bar{p}/p$, $\bar{\Lambda}$/$\Lambda$, $\bar{\Xi}$/$\Xi$, $\bar{\Omega}$/$\Omega$, $k^-$/$\pi^-$, $\bar{p}$/$\pi^-$, $\Lambda$/$\pi^-$, $\Xi$/$\pi^-$ and $\Omega$/$\pi^-$ except for $200$ GeV, we also use $(\Omega+\bar{\Omega})/\pi^-$ instead of $\Omega$/$\pi^-$, i.e. $11$ particle ratios.  At $39$, $11.5$ and $7.7$ GeV, $\pi^-$/$\pi^+$, $k^-$/$k^+$, $\bar{p}$/p, $\bar{\Lambda}$/$\Lambda$, $\bar{\Xi}$/$\Xi$, $\bar{\Omega}$/$\Omega$, $k^-$/$\pi^-$, $\bar{p}$/$\pi^-$, $\Lambda$/$\pi^-$ and $\bar{\Xi}$/$\pi^-$, i.e. $10$ particle ratios, are used.  Further details about the given data sets at various STAR energies are in order.    

\begin{description}
\item{\bf At $200~$GeV}, we use yields of pions, kaons, (anti)protons~\cite{STAR2}, $\Lambda$, $\bar{\Lambda}$ and multi-strange baryons~\cite{STAR3,STAR4} measured at mid-rapidity measured in  the STAR experiment at centrality $0-5\%$, except the $\Omega$/$\bar{\Omega}$ ratios are measured at $0-20\%$~\cite{STAR4}. The measured pions spectra are corrected for feed-down from weak decays as well as $\Lambda$ ($\bar{\Lambda}$) are corrected for feed-down from weak decays of multi-strange baryons. The comparison between the experimental (symbols) and calculated ratios (horizontal lines) is shown in left panel of Fig. \ref{fig200_130}. For comparison with other ratios and even better appearance, some ratios are scaled to avoid log plot. 

\begin{figure}[htb]
\centering{
\includegraphics[width=8.5cm]{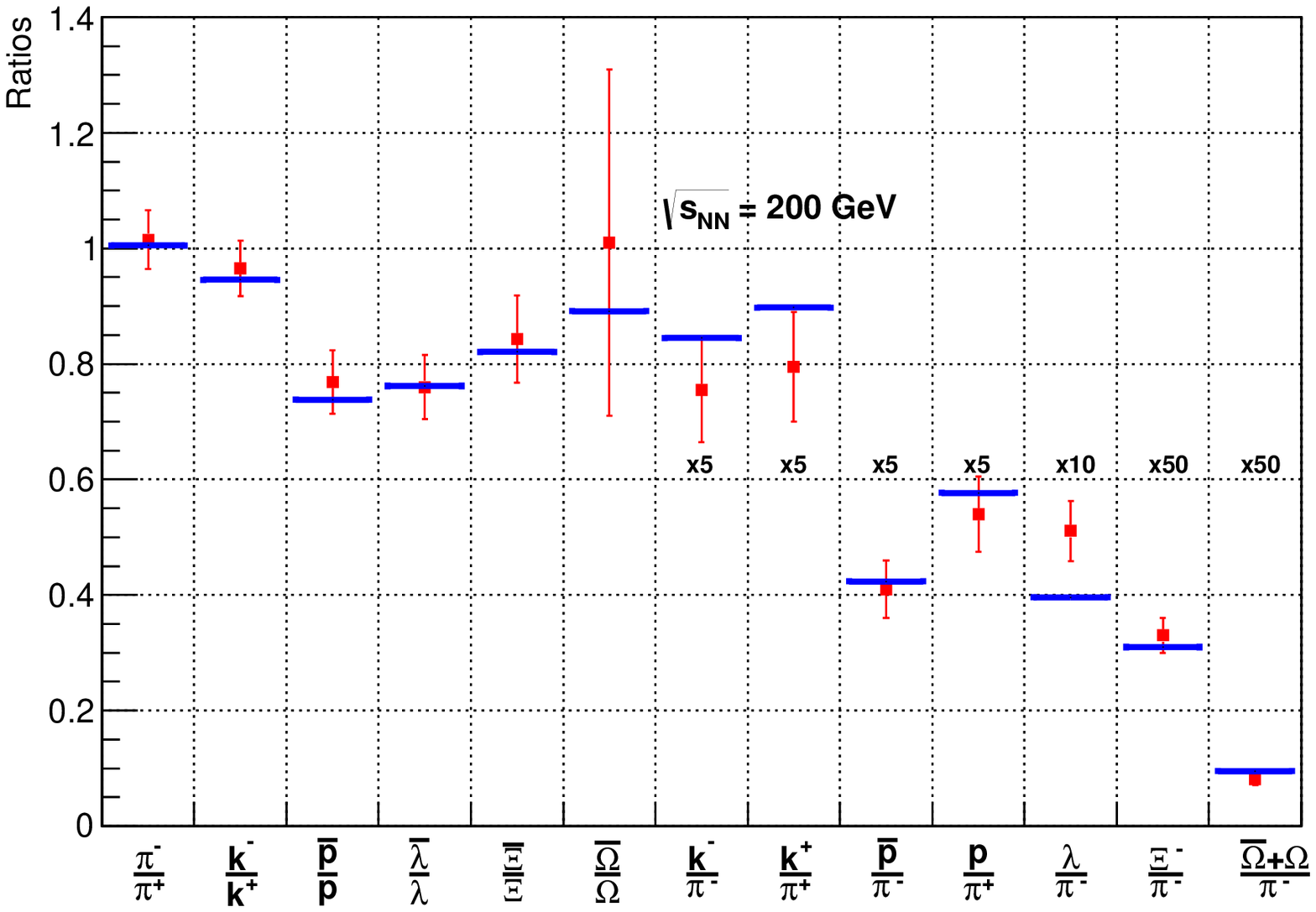}
\includegraphics[width=8.5cm]{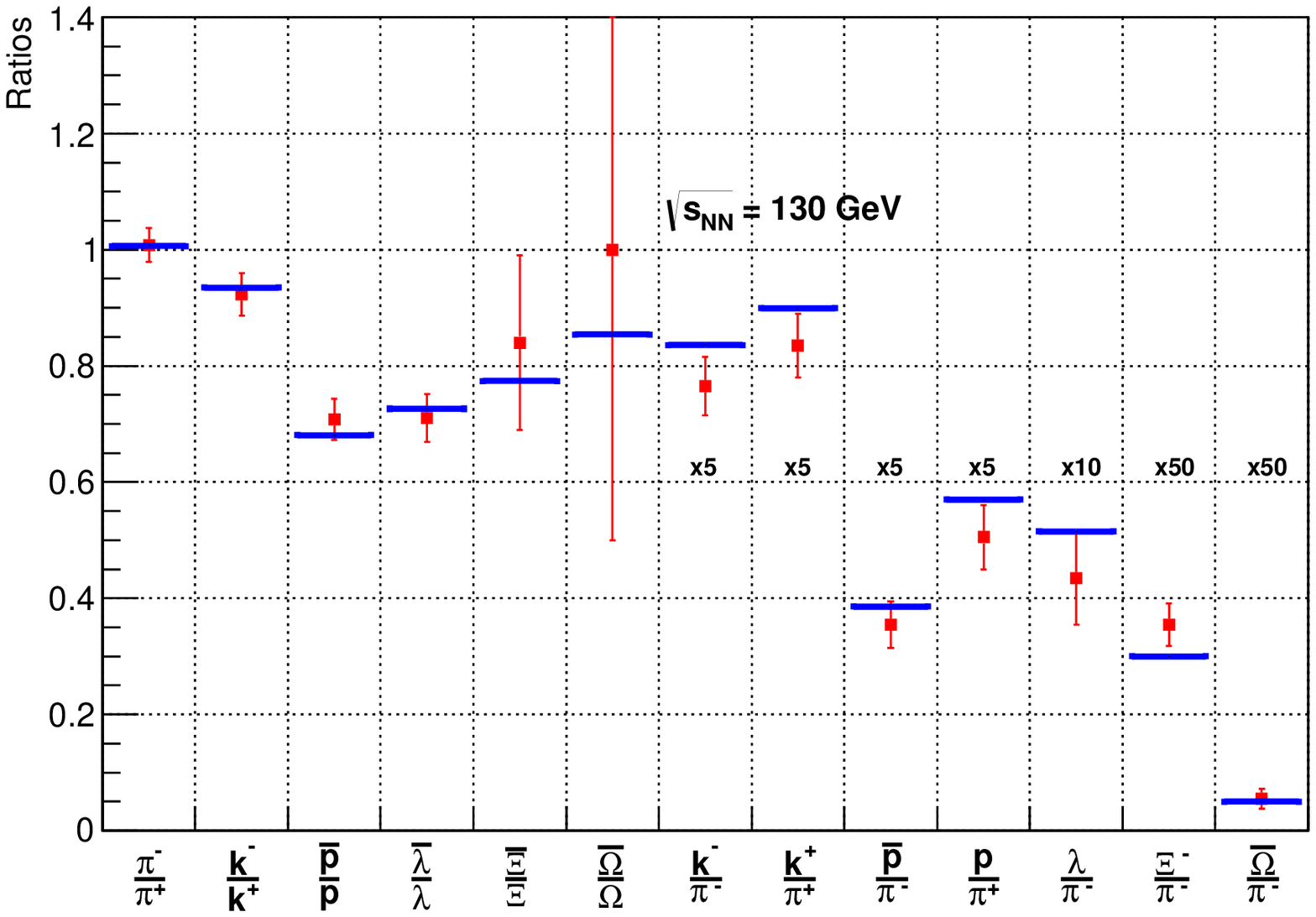}
\caption{Left panel: the experimental particle ratios (symbols) ~\cite{STAR2,STAR3,STAR4} are compared to the HRG calculations (horizontal lines) at $200~$GeV,  where HRG calculations was preformed at $T_{ch}$ and $\mu_b$ parameters which assure minimum $\chi^2$ per degrees-of-freedom. For comparison with other ratios and better appearance, some ratios are scaled (scaling factor are given) so that we avoid log plotting. Right panel shows the same as in left panel but at $130~$GeV, where the experimental particle ratios are taken from Refs.~\cite{STAR2,STAR5,STAR4,STAR6} 
 \label{fig200_130} }
}
\end{figure}

\item{\bf At $130~$GeV}, we use yields of pions, kaons, (anti)protons~\cite{STAR2}, $\Lambda$, $\bar{\Lambda}$~\cite{STAR5} and multi-strange baryons~\cite{STAR4,STAR6} measured at mid-rapidity measured in the STAR experiment at centrality $0-5\%$ except the multi-strange baryons have been measured at $0-20\%$. The measured pions spectra are corrected for feed-down from weak decays. The results comparing STAR with HRG ratios are shown in right panel of Fig. \ref{fig200_130}.

\item{\bf At $62.4~$GeV}, we use yields of pions, kaons, (anti)protons~\cite{STAR2}, $\Lambda$, $\bar{\Lambda}$ and multi-strange baryons~\cite{STAR4} measured at mid-rapidity measured in the STAR experiment at centrality $0-5\%$ except the  multi-strange baryons are measured at $0-20\%$. The measured pion spectra are corrected for feed-down from weak decays as well as $\Lambda$ ($\bar{\Lambda}$) are corrected for feed-down from weak decays of $\Xi$. The comparison between STAR and HRG results is shown in left panel of Fig. \ref{fig62_39}.
   
\begin{figure}[htb]
\centering{
\includegraphics[width=8.5cm]{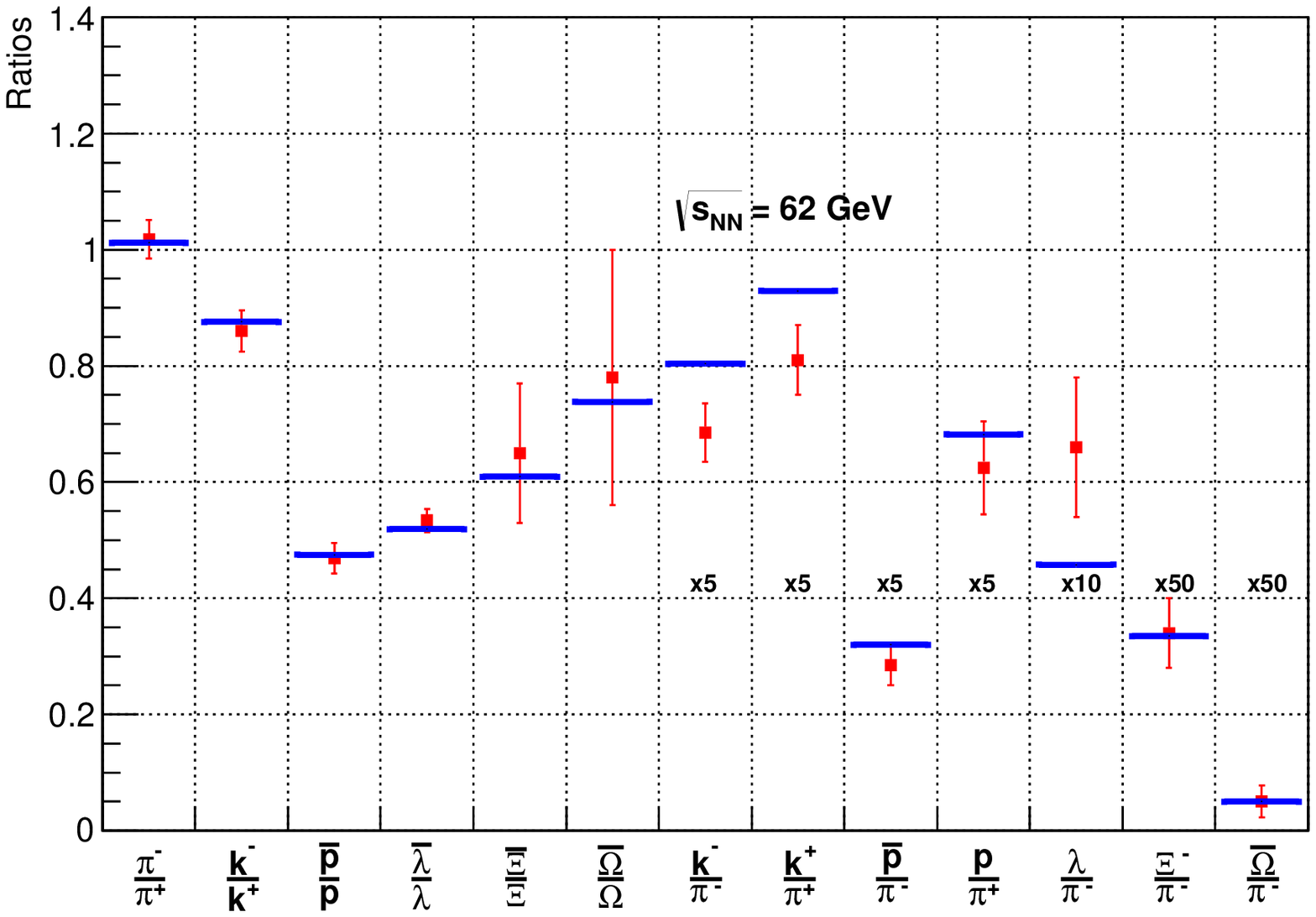}
\includegraphics[width=8.5cm]{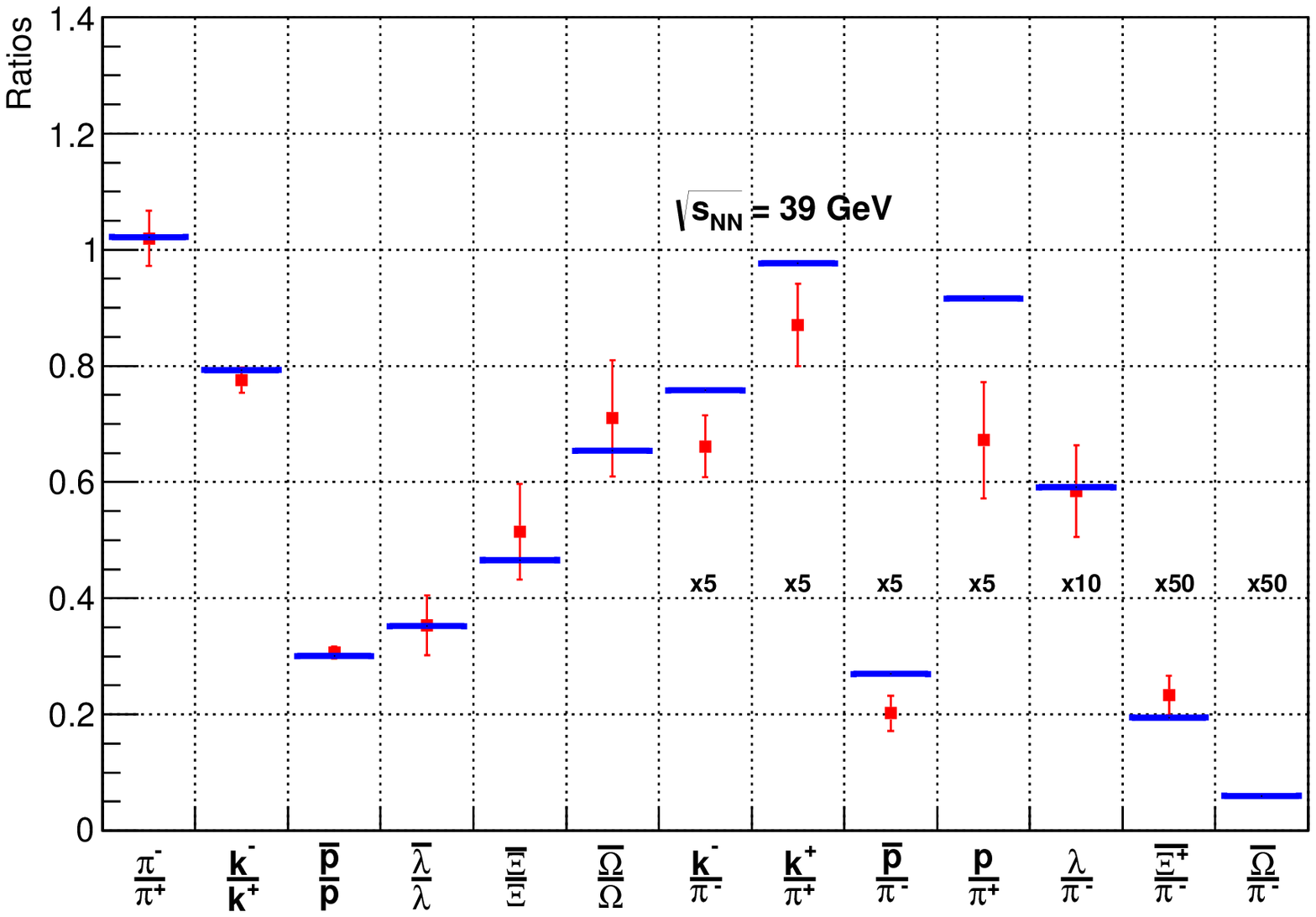}
\caption{The same as in left panel of Fig. \ref{fig200_130} but at $62.4~$GeV (left panel), where the experimental particle ratios are taken from Refs.~\cite{STAR2,STAR4} and at $39~$GeV (right panel), where the experimental particle ratios are taken from Refs.~\cite{STAR7a,STAR7b,STAR8,privateCommunication,calculated}.  \label{fig62_39} }
}
\end{figure}

\item{\bf At $39$, $11.5$ and $7.7~$GeV}, we use for yields of pions, kaons, (anti)protons~\cite{STAR7a,STAR7b}, $\Lambda$, $\bar{\Lambda}$,  $\Xi$ and $\bar{\Xi}$~\cite{STAR8} measured at mid-rapidity measured in the STAR experiment at centrality $0-5\%$. The $\Omega$/$\bar{\Omega}$ ratios at centrality $0-5\%$, $0-20\%$ and $0-60\%$ for $39$, $11.5$ and $7.7~$GeV, respectively, are taken from Ref.~\cite{privateCommunication}. The measured pion spectra have been corrected for feed-down from weak decays as well as $\Lambda$ ($\bar{\Lambda}$) for the feed-down contributions from $\Xi$ weak decay. The analysis includes $10$ independent ratios where $\Omega/\pi$ is excluded. The results are shown in right panel of Fig. \ref{fig62_39} and Fig. \ref{fig11_7}.

\begin{figure}[htb]
\centering{
\includegraphics[width=8.5cm]{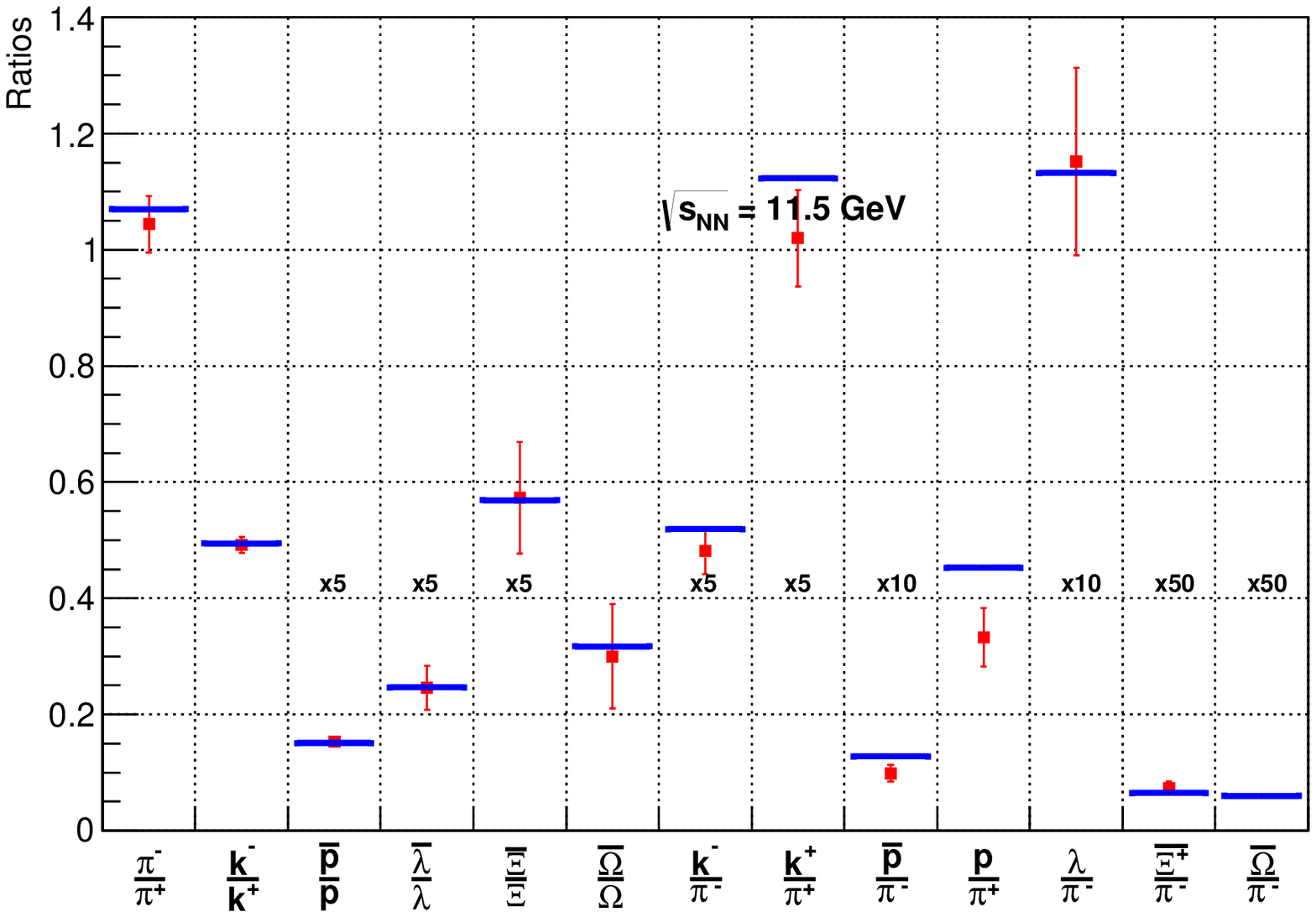}
\includegraphics[width=8.5cm]{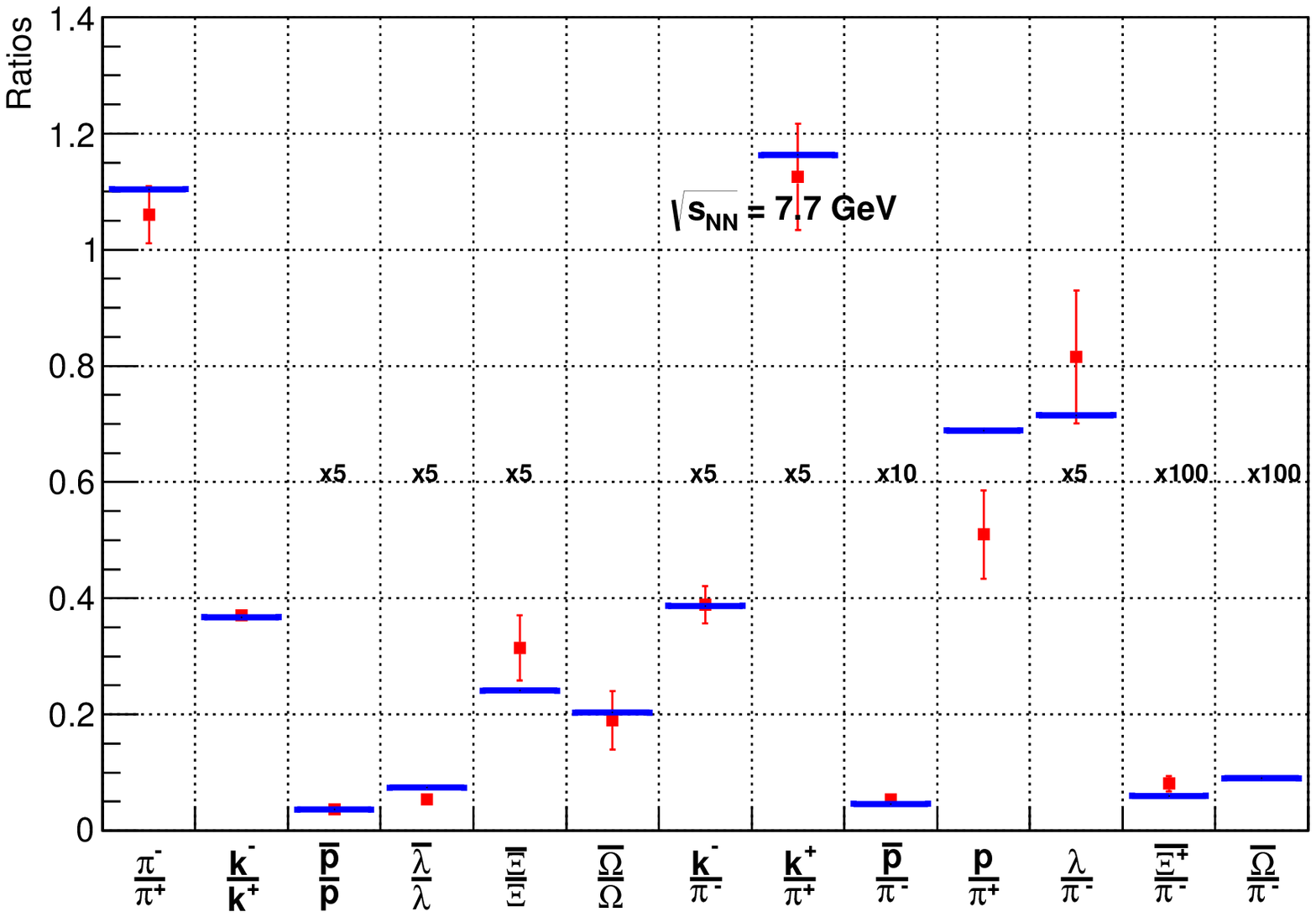}
\caption{The same as in left panel Fig. \ref{fig200_130} but at $11.5~$ (left panel) and $7.7~$GeV (right panel) where experimental data are taken from Refs.~\cite{STAR7a,STAR7b,STAR8,privateCommunication,calculated}. \label{fig11_7}}
}
\end{figure}

\end{description}

\begin{table}[htb]
%\centering
\begin{tabular}{|c|  | c    | c    | c    || c   | c    | c || c |    c |    c    || c   | c    | c |}     
 % centered columns (6 columns)
\hline
 & \multicolumn{6}{|c||}{$r_m=r_b=0.0~ $fm} &\multicolumn{6}{|c|}{$r_m=r_b=0.3~$fm}\\
\hline     %inserts double horizontal lines
 $\sqrt{s_{NN}}$ [GeV] & $T_{ch}$  & $\mu_b$ & $\chi^2/dof$ & $T_{ch}$ & $\mu_b$ &  $q^2$  & $T_{ch}$  & $\mu_b$ & $\chi^2/dof$ & $T_{ch}$ & $\mu_b$ &  $q^2$ \\ [0.5ex] % inserts table      %heading
\hline       % inserts single horizontal line
200 & 159.5 & 27.5 & 9.24/9 & 161.5  & 30 & 0.137 & 159.5 & 27.5 & 9.24/9 & 161.5  & 29.5 & 0.137 \\   % inserting body of the table
130 & 157.5  & 34 & 6.86/9 & 160  & 32 & 0.108  &157  & 34 & 7.07/9 & 159.5  & 32 & 0.109 \\
62.4 & 157.5 & 66.5 & 10.17/9 & 161.5  & 73 & 0.185  & 157 & 66 & 10.51/9 & 161.5  & 72.5 & 0.185\\
39 & 160.5 & 110.5 & 10.69/8 & 163.5  & 113.5 & 0.122 & 160.5 & 110.5 & 11.03/8 & 163  & 112.5 & 0.124  \\
11.5 & 153 & 308 & 6.06/8 & 153.5  & 312 & 0.072 & 153 & 308 & 6.28/8 & 153  & 310 & 0.073 \\ 
7.7 & 145.5 & 410.5 & 15.21/8 & 149  & 412.5 &  0.265 & 145 & 409 & 15.07/8 & 148.5 & 419 & 0.263   \\ [1ex]
 % [1ex] adds vertical space
\hline
 %inserts single line
\end{tabular}
\caption{At STAR energies, the freeze-out parameters, $T_{ch}$ and $\mu_b$, are estimated from $\chi^2$ and $q^2$ fitting approaches assuming point-like and single hard-core for ( $r_m=r_b=0.3~$fm) constituents of the HRG model. In both fittings, the degrees-of-freedom (dof) are given. \label{tab:1}}
\end{table}

The freeze-out parameters, $T_{ch}$ and $\mu_b$, are estimated from $\chi^2$ and $q^2$ fitting approaches assuming point-like and finite hard-core (single hard-core radius, $r_m=r_b=0.3~$fm) constituents of the HRG model. They are listed out in Tab. \ref{tab:1}. Few remarks are now in order. 
\begin{itemize}
\item The difference between the freeze-out parameters extracted from the HRG model with vanishing and finite hard-core is too small. In light of this, we believe that the results at $r_m=r_b=0.0~$fm are realistic. 
\item The same behavior is also observed in both $\chi^2$ and $q^2$ (almost no difference). The parameters extracted using $q^2$ are a little bit larger than the ones using $\chi^2/dof$ reflecting the accuracy of extracted parameters on $\sqrt{s_{NN}}$. 
\item It is obvious that the values of $\chi^2/dof$ are close to unity, except at $7.7~$GeV. The reason may be the need to assume a degree of non-equilibrium, especially at this relative low energy. This will be analysed in a future work.
\item The main difference between both fitting methods is that $\chi^2$ takes into consideration the uncertainty of the experiment. In $\chi^2/dof$, not only the ratios which are sensitive to $T_{ch}$ and $\mu_b$ will have the upper hand in determining the extracted parameters, but also the ratios with higher measured accuracy. On other hand, $q^2$ is designed to reflect the deviation. Therefore, $\chi^2/dof$ seems to be more suitable to be implemented.
\end{itemize}
 Depending on the previous notes, we focus on the results of $\chi^2/dof$ method taking $\sim 1\, \sigma$-error at $r_m=r_b=0.0~$fm. The final results are summarized in Tab. \ref{tab:2}.

\begin{table}[htb]
\centering
\begin{tabular}{|c|  | c  | c |   }      % centered columns (6 columns)
\hline\hline     %inserts double horizontal lines
$\sqrt{s_{NN}}~$[GeV] & $T_{ch}$ [MeV] & $\mu_b$ [MeV] \\ [0.5ex] % inserts table      %heading
\hline       % inserts single horizontal line
200 & $159.5 \pm 2   $ & $27.5 \pm 4  $   \\   % inserting body of the table
130 & $157.5 \pm 4  $ & $ 34\pm 3.5  $   \\
62.4 & $157.5 \pm 3.5  $ & $66\pm 4  $   \\
39 & $160.5 \pm 3.5  $ & $110.5\pm 4.5  $   \\
11.5 & $153 \pm 1.5  $ & $308\pm 2.5  $   \\ 
7.7 & $145.5 \pm 1  $ & $410.5\pm 4  $  \\ [1ex]
 % [1ex] adds vertical space
\hline\hline
 %inserts single line
\end{tabular}
\caption{The freeze-out parameters estimated from $\chi^2/dof$ best fit of various experimental particle ratios compared with the HRG calculations with $r_m=r_b=0.0~$ fm.}
\label{tab:2}
 % is used to refer this table in the text
\end{table}

\section{Results and Discussion}
\label{sec:res}

\subsection{Effects of Excluded-Volume Correction(EVC)} 

The differences between extracted thermal parameters using finite hard-core for hadrons ($r_m=r_b=0.3~$fm) and vanishing one ($r_m=r_b=0.0~$fm) is nearly negligible. Actually, if one uses Maxwell-Boltzmann statistics, the number density of $i$-th particle will be suppressed by $R_i$, Eq. (\ref{eq:R}),~\cite{EVC} 
\bea \label{eq:n}
n^{excl}_i(T,\mu_i) &=& R_i\, n^{id}_i(T,\mu_i),
\eea
where
\bea \label{eq:R}
R_i(T,\mu_i,\upsilon)  &=& \frac{\exp(\frac{-\upsilon_i\, p^{excl}}{T})}{1+\sum_i \upsilon_i\, n^{id}_i(T,\tilde{\mu_i}) }.
\eea 
Then, the ratios between two particle species at finite excluded volume will be 
\bea \label{eq:nn}
\frac{n^{excl}_i(T,\mu_i)}{n^{excl}_j(T,\mu_i)} &=& \frac{R_i\, n^{id}_i(T,\mu_i)}{R_j\, n^{id}_j(T,\mu_i)}. 
\eea               
Therefore, the ratios should not be effected, when $r_m=r_b$. In the present paper, we use quantum statistics which makes a very small shift in the extracted thermal parameters. Both sides of Eq. (\ref{eq:nn}) do not become exactly equal, but still represent a good approximation. 

As given in Tab. \ref{tab:1}, although the effect of EVC on the extracted thermal parameters can be completely ignored,  when $r_m = r_b$, the effect would increase when different hard-core radii for mesons and baryons ($r_m \neq r_b$) are assumed. The effect of EVC, when $r_m = r_b$ appears in thermodynamic quantities, like energy, entropy and number densities, as seen in Eq. (\ref{eq:n}).

\subsection{Effects of Weak Decay from Feed-Down} 

\begin{figure}[hbt]
\centering{
\includegraphics[width=8.5cm]{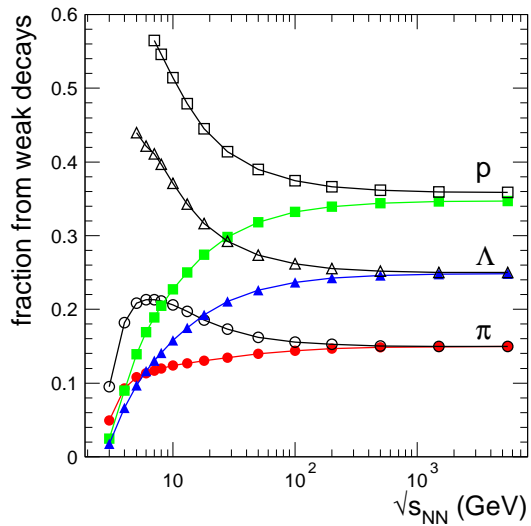}
\caption{The energy dependence of the fraction of total hadron yields originating from weak decays. The full symbols are for particles ($\pi^+$, p, $\Lambda$), while the open ones for antiparticles ($\pi^-$, p, $\bar{\Lambda}$). Different asymptotic values are connected with different particle species. The contributions of weak-decay from feed-down of particle and antiparticles coincide in the asymptotic region. Graph taken from Ref. \cite{Andronic}. \label{weak} }
}
\end{figure}

The contribution of feed-down from weak decay has been studied in Ref.~\cite{Andronic}. The results are given in Fig. \ref{weak}. The energy dependence of the contribution of weak decays to yields of pions, protons and hyperons is calculated with parametrizations of the freeze-out parameters, $T_{ch}$ and $\mu_b$. Concretely, the figure shows the fraction of the total yield of these particles which are originating from weak decays. We notice that this quantity reaches asymptotic values of $15\%$, $25\%$ and $35\%$ for pion, lambda and proton, respectively. The quantity should be significantly large for antiparticles at high energies~\cite{Andronic}. We find that assuming different contributions from feed-down from weak decay leads to different extracted thermal parameters~\cite{Andronic,MichalecPh.D}. In light of this, the inclusion of the feed-down contributions in the HRG model should be implemented. 

\subsection{Effects of Including New Resonances} 

The discovery of new hadron resonances apparently affects the values of extracted freeze-out parameters $T_{ch}$ and $\mu_b$~\cite{Andronic2009}. Furthermore, the particle $\sigma$ labelled as $f_0(600)$ would have an effect on the $k^+/\pi^+$ ratio, the so-called {\it ''horn problem''}~\cite{Andronic2009,Satarov}. Therefore, the inclusion of $\sigma$ would lead to better fitting, especially at high energies. 
\begin{itemize}
\item  We have repeated the analysis without $\sigma$ aiming to estimate its effect on the freeze-out parameters  at $200~$ and $130~$GeV. We get $\mu_b=27$, and $T_{ch}=157~$ MeV at $200~$GeV and  $\mu_b=33.5~$, $T_{ch}=154.5~$ MeV at $130~$GeV. Comparing these with the corresponding ones in Tab. \ref{tab:2} ($\mu_b=27.5$, $T_{ch}=159.5~$MeV and $\mu_b=34$, $T_{ch}=157.5~$MeV, respectively) shows that improvements through including $\sigma$ lay within the statistical errors.
\item We repeat the analysis at $200~$ and 130 GeV, when resonance masses $ \le 2~ GeV$.  The same results were obtained.
\end{itemize}
We conclude that the discovery of new particles (especially the light ones like $\sigma$) would affect the extracted freeze-out parameters.

\subsection{Energy Dependence of the Freeze-out Parameters} 

\begin{figure}[htb]
\centering{
\includegraphics[width=8.5cm,height=8.cm]{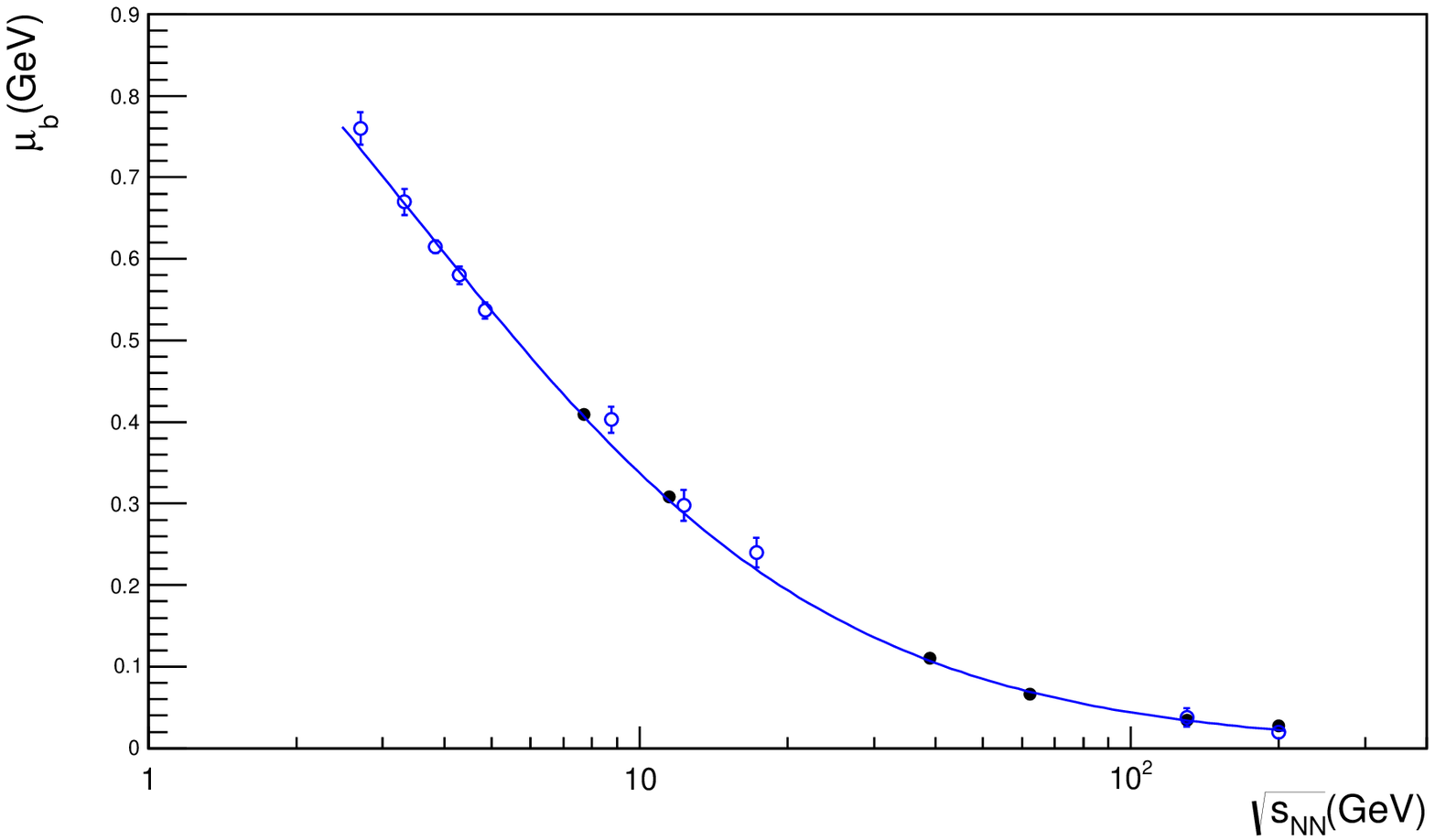}
\includegraphics[width=8.5cm,height=8.cm]{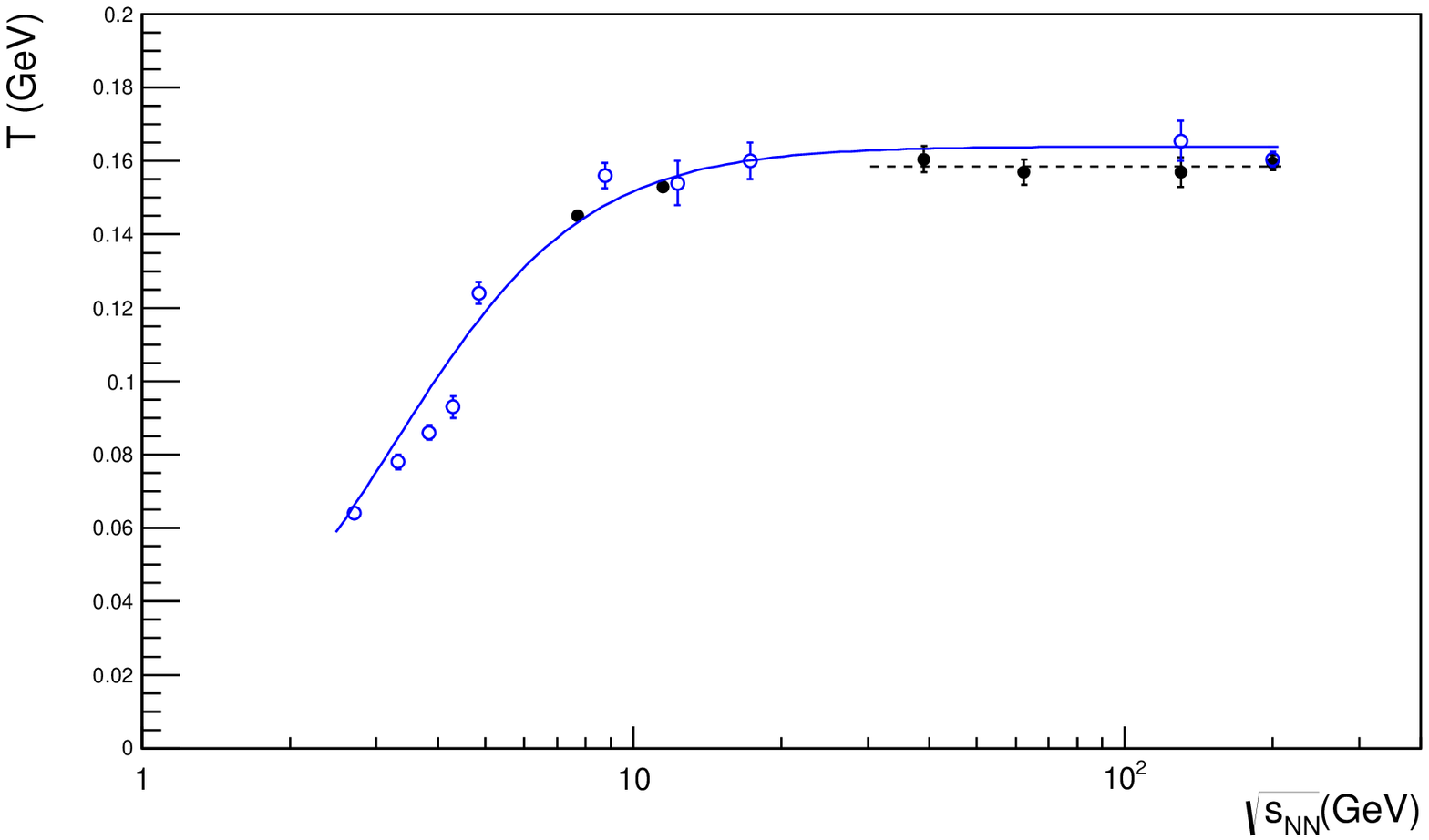}
\caption{The energy dependent of the extracted parameters $\mu_b$ (left panel) and  $T_{ch}$ (right panel). The open symbols represent results of Ref.~\cite{Andronic}. The closed symbols give the results of this paper. The solid curves stand for  Eqs. (\ref{eq:MuB}) and Eq(\ref{eq:T}), respectively. Dashed line gives the best fit of four points. The whole data set seems not enough to construct parametrization.
 \label{TmuBsqrts} }
 }
\end{figure}

Estimating different freeze-out parameters at different energies raises the question about the systematic energy dependence. In Fig. \ref{TmuBsqrts}, we compare our results (closed symbols) with previous studies (empty symbols)~\cite{Andronic,Andronic2009}. The dependence of  $\mu_b$ on $\sqrt{s_{NN}}$, left panel, can be parametrized \cite{Andronic,Andronic2009} as
\bea 
\mu_b &=& \frac{a}{1+b\sqrt{s_{NN}}}, \label{eq:MuB}
\eea 
where both $\mu_b$ and $\sqrt{s_{NN}}$ are given in GeV units.  According to the present analysis $a=1.245 \pm 0.094~$GeV and $b=0.264 \pm 0.028~$GeV$^{-1}$. The solid curve in left panel on Fig. \ref{TmuBsqrts} represents these values. According to Ref. \cite{Andronic,Andronic2009}, $a=1.303 \pm 0.120~$GeV and $b=0.286 \pm 0.049~$GeV$^{-1}$.

Also, the dependence of  $T_{ch}$ on $\sqrt{s_{NN}}$ can be parametrized~\cite{Andronic2009} 
\bea 
T_{ch} &=& T_{lim}\left(\frac{1}{1+\exp\left[\frac{1.172-\ln(\sqrt{s_{NN}})}{0.45}\right]}\right), \label{eq:T}
\eea 
where $\sqrt{s_{NN}}$ are given in GeV. The limiting temperature reads, $T_{lim}=164~$MeV. This value is slightly higher than the one obtained in Ref.~\cite{Andronic}, $T_{lim}=161\pm4~$MeV. The fit of just four point at the highest energies is given by the dashed curve. The whole data set consisting of six points would not be enough to construct parametrization as given in Eq. (\ref{eq:T}). The reason for this is the statistical accuracy.

STAR BES II is designed to include further energies, such as $17$ and $2.2~$GeV. Also, NICA and FAIR facilities would cover other energies. In light of this, the potential to extend this analysis and enrich it with new measurements is very likely. Currently, we restrict the conclusion to STAR BES I and older energies (six freeze-out parameters). 

\subsection{Regularities in resulting Freeze-Out Parameters}

\begin{figure}[htb]
\centering{
\includegraphics[width=10.0cm,height=8.cm]{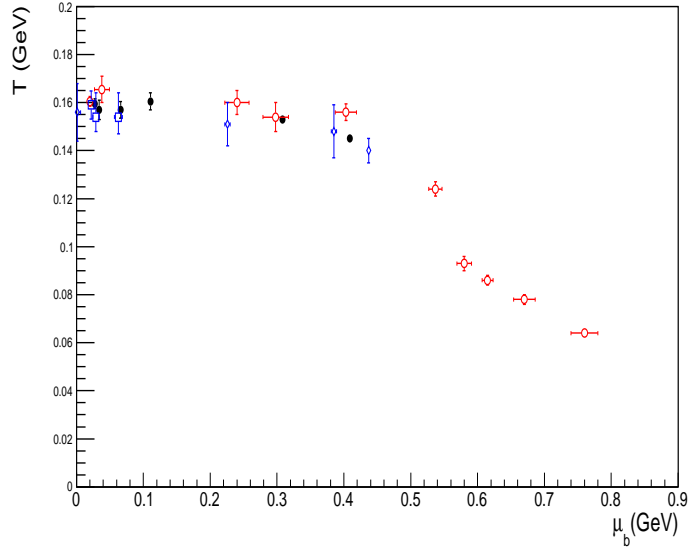}
\caption{The regularities in the extracted thermal parameters are shown in the freeze-out diagram. The open circle symbols are results from Ref.~\cite{Andronic}, the open diamond symbols the results~\cite{SHMUrQM}, the open square are STAR results (at 0-5\% centrality)~\cite{STAR2} and the closed symbols are the new result of our analysis. Within the range covered by the present analysis, the agreement is convincing.
 \label{fig:TmuB} }
 }
\end{figure}

Fig. \ref{fig:TmuB} shows the dependence of $T_{ch}$ on $\mu_b$, the freeze-out diagram. The results of Ref.~\cite{Andronic,SHMUrQM} (open symbols) are compared with that of present analysis (closed symbols). Also, the results taken from Ref.~\cite{SHMUrQM} were obtained using statistical hadronization model with partial chemical equilibrium ($\gamma_s \ne 1$). The STAR points~\cite{STAR2} were obtained using the particle ratios, in which strange baryons ratios were not included. The comparison between lattice QCD chiral phase transition temperature and the HRG model seems to confirm closeness of chemical freeze-out parameters and the QCD transition line at relative low $\mu_b$. With closeness we mean that the temperature difference is relatively small.

As given  in Fig. \ref{TmuBsqrts}, the resulting freeze-out temperature is approximately constant. Over a range of $\sim400~$MeV chemical potential, the value of $T_{ch}$ decreases of $\sim14~$MeV. This would reflect that QGP might be close to the hadronic phase, especially at energies $\ge 39~$GeV. Also, we note that $T_{ch}$ gets relative low values at $11.5~$GeV. This would indicate a critical point.  

We have so-far estimated just six points, i.e. a small range of temperatures. However, we believe that the new results can give a better estimation for $T_{lim}$, because of including relative high-energy measurements, $200$, $130$, $62.4$ and $39~$GeV. Taking the average of these four points, then $T_{lim}=158.5 \pm 3~$MeV (dashed line in right panel).

\subsection{Comparison with Universal Freeze-Out Conditions}

Starting from phenomenological observations at SchwerIonen-Synchrotron (SIS) energy, it was found that the averaged energy per averaged particle $\epsilon/n \approx 1~$GeV \cite{jeanRedlich}, where Boltzmann approximations are applied in estimating $\epsilon/n$. This constant ratio is assumed to describe the whole $T_{ch}-\mu_b$ diagram. The second criterion assumes that total baryon number density $n_b+n_{\bar{b}} \approx 0.12~$fm$^{-3}$ \cite{nb01}. In framework of percolation theory~\cite{percl}, a third criterion has been suggested. As shown in Fig. 2 of Ref. \cite{Tawfik:2005qn}, the last two criteria seem to give almost identical results. A fourth criterion based on lattice QCD simulations was introduced in Ref.  \cite{Tawfik:2005qn,Tawfik:2004ss}. Accordingly, the entropy normalized to cubic temperature is assumed to remain constant over the whole range of baryon chemical potentials, which is related to $\sqrt{s_{NN}}$ \cite{jean2006}.

In framework of hadron resonance gas, the thermodynamic quantities deriving the chemical freeze-out are deduced \cite{Tawfik:2005qn,Tawfik:2004ss}. The motivation of suggesting constant normalized entropy is the comparison to the lattice QCD simulations with two and three flavors. We simply found the $s/T^3=5$ for two flavors and $s/T^3=7$ for three flavors. Furthermore, we confront the HRG results to the experimental estimation for the freeze-out parameters, $T$ and $\mu_b$. In rest frame of produced particle, the hadronic matter can be determined by constant degrees of freedom, for instance, $s/T^3 (4/\pi^2)=const$ \cite{Tawfik:2005qn,Tawfik:2004ss}. The chemical freeze-out can be related to particle creation. Therefore, the abundances of different particle species should be controlled by $\mu_b$, which obviously depends on $T$. With the beam energy, $T$ is increasing, while the baryon densities (or $\mu_b$) at mid-rapidity is decreasing. 

A fifth criterion used higher order moments of particle multiplicity~\cite{Tawfik:2013dba,HM_FO} assumed that the freeze-out parameters are defined, when $\kappa\, \sigma^2$ vanishes \cite{Tawfik:2013dba}
\bea \label{eq:cond1}
\frac{\int_0^{\infty} k^2\, dk \left[\cosh(\frac{\varepsilon-\mu}{T})\pm2\right] \left[\exp(\frac{\varepsilon-\mu}{2T})\mp\exp(\frac{\mu-\varepsilon}{2 T})\right]^4}{\int_0^{\infty} k^2\, dk \left[\exp(\frac{\varepsilon-\mu}{2T})\mp\exp(\frac{\mu-\varepsilon}{2T})\right]^2} &=&\frac{3}{4}.
\eea
or
\bea
\label{eq:cond2}
\left\langle\cosh\left(\frac{\varepsilon-\mu}{T}\right)\pm 2\right\rangle &=&\frac{3}{4}.
%\\ \left\langle\cosh\left(\frac{\varepsilon-\mu}{T}\right)\right\rangle &=&\frac{3}{4} \mp 2
\eea
It should be noted that the quantity $\varepsilon$ varies from resonance to another. 

The sixth criterion assumes the trace anomaly should remain unchanged over the freeze-out diagram \cite{Tawfik:2013eua}
\bea \label{eq:Icls}
\frac{I(T,\mu_b)}{T^4} &=& \frac{I(T)}{T^4} - \frac{\chi_2(T,\mu_b)}{T^2}\, \frac{\mu_b^3}{2 T} + \frac{g}{8 \pi^2}\; e^{\mu_b/T}\; \mu_b^2 \left(\frac{m}{T}\right)^3  \left[K_1\left(\frac{m}{T}\right) + K_3\left(\frac{m}{T}\right)\right].
\eea
Assuming that $I(T,\mu_b)/T^4=7/2$ at vanishing and finite $\mu_b$, then
\bea
\frac{1}{2}  &=&  T \left(\frac{m}{T}\right) \, K_2\left(\frac{m}{T}\right) \, \left[K_1\left(\frac{m}{T}\right) + K_3\left(\frac{m}{T}\right)\right].
\eea

\begin{figure}[htb]
\centering{
\includegraphics[width=10.0cm,angle=-90]{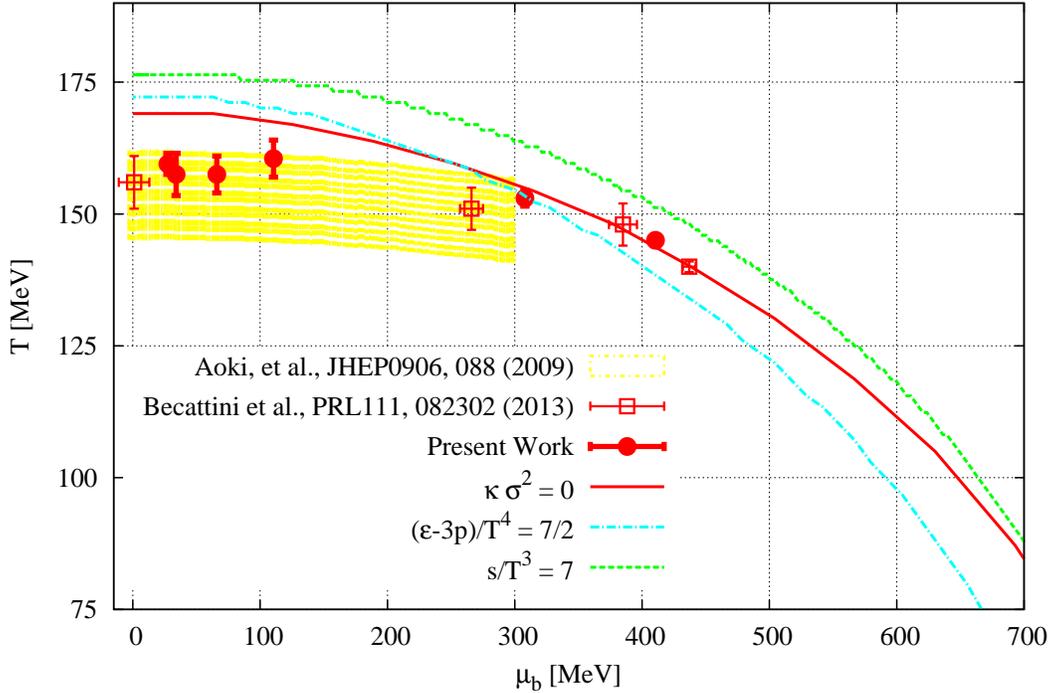}
\caption{The freeze-out parameters are compared with lattice QCD calculations \cite{Karsch,LQCD} (band) and Becattini {\it et al.} (open symbols) \cite{SHMUrQM}. The three curves represent three universal conditions suggested to describe the regularities in the freeze-out parameters. The solid curve represents vanishing $\kappa\, \sigma^2$ \cite{Tawfik:2013dba}. The dashed and dash-dotted curves give the regularities according to $s/T^3=7$  \cite{Tawfik:2005qn,Tawfik:2004ss} and $(\epsilon-3P)/T^4=7/2$ \cite{Tawfik:2013eua}, respectively.
 \label{fig:TmuB_comp} }
 }
\end{figure}

Fig. \ref{fig:TmuB_comp} depicts the freeze-out parameters and compares them with the  lattice QCD calculations \cite{Karsch,LQCD} (band) and Becattini {\it et al.} (open symbols) \cite{SHMUrQM}. The solid curve represents vanishing $\kappa\, \sigma^2$ \cite{Tawfik:2013dba}, while dashed and dash-dotted curves give the regularities according to $s/T^3=7$  \cite{Tawfik:2005qn,Tawfik:2004ss} and $(\epsilon-3P)/T^4=7/2$ \cite{Tawfik:2013eua}, respectively. The agreement with the three criteria is not convincing. Vanishing $\kappa\, \sigma^2$ \cite{Tawfik:2013dba} is relatively close to the freeze-out parameters, especially at large chemical potentials. 

\section{Conclusions and Outlook}
\label{cons}

Over the last three decades, heavy-ion experiments at energies ranging from SIS and LHC (utilizing different detectors and thus having different accuracies) have been carried out. The regularities in the produced particles at different energies haven been analysed, for instance, the freeze-out temperature and baryon chemical potential can be deduced from statistical fits of experimentally-estimated particle ratios to thermal model calculations assuming chemical equilibrium.

The systematic study of the freeze-out parameters dates back to 2006 \cite{Andronic,jean2006}. Since that time, LHC was being commissioned and STAR launched its BES-I. STAR BES-I enables measurements to be made at energies ranging from SPS to top RHIC with the same detector, offers same uniform acceptance at each energy value and finally, not only the statistics gets better due to the higher acceptance of the STAR detector, but also, there is cleaner and more extensive PID capabilities. This allows not only to repeat the SPS measurements in much finer details but also to enhance them through differential and multiple measures. The latter is very essential, since integrating measures because of small statistics or limited acceptance is connected with lost in valuable information. Because of the so-called anomaly in proton, and antiproton and proton-to-pion ratios measured by the ALICE experiment ~\cite{ALICE,Petran,SHMUrQM}, we did not include LHC in the present analysis. Also,  because of the irregularities registered in the SPS measurements, which led to large differences in the extracted freeze-out parameters, for instance, NA49, NA44, and NA57 measurements at $17.3~$GeV ~\cite{Andronic}, we exclude SPS, as well. 

The present analysis includes $11$ (occasionally $10$) independent particle-ratios. Additional particle-ratios were included, as soon as these are available. The number of particle ratios is kept unchanged at all energies. The best fits are the ones assuring the simultaneous regeneration of the $11$ independent particle-ratios. the fits to individual particle ratios are not taken into consideration. Furthermore, this rule is respected at all energies. Another rule governing the present analysis is that the hadron ratios - with corresponding errors - are taken from STAR BES-I. In some cases, the ratios - and their corresponding errors - have been calculated from the published yields, as no experimentally-estimated particle ratios were available.   

The freeze-out parameters, $T_{ch}$ and $\mu_b$, are estimated from $\chi^2$ and $q^2$ fitting approaches assuming point-like and finite hard-core (single hard-core radius, $r_m=r_b=0.3~$fm) constituents of the HRG model. We find that the difference between the freeze-out parameters extracted from the HRG model with vanishing and finite hard-core is too small. The extracted parameters extracted using $q^2$ are a little bit larger than the ones using $\chi^2/dof$ reflecting the accuracy of extracted parameters on $\sqrt{s_{NN}}$. The main difference between both fitting methods is that $\chi^2$ takes into consideration the uncertainty of the experiment. In $\chi^2/dof$, not only the ratios which are sensitive to $T_{ch}$ and $\mu_b$ will have the upper hand in determining the extracted parameters, but also the ratios with higher measured accuracy. On other hand, $q^2$ is designed to reflect the deviation. Therefore, $\chi^2/dof$ seems to be more suitable to be implemented.

We conclude that the particle ratios are not affected when using finite $r_m=r_b$. The EVC makes an equal shift in the yields, and simultaneously in the thermodynamic quantities. The importance of including a clear specific weak decay contribution in HRG appears, for instance, in the difference between the extracted freeze-out parameters at $130~$Gev and the values given in Ref.~\cite{Andronic}. We believe that the difference is due to the PHENIX data taken into consideration in Ref.~\cite{Andronic}, in which no clear specific estimation for the contribution of weak decay~\cite{PHENIX} in pion spectra has been performed. The relations between $T_{ch}$ and $\mu_b$ vs $\sqrt{s_{NN}}$ are sensitive to the rapidity cut in the experiment. The particle ratios are affected by the rapidity~\cite{BRAHMS}. Therefore, the obtained parameters depend on the rapidity cut and centrality, as well as. 

The discovery of new resonances, especially with low masses, will affect the extracted freeze-out parameters. This remains a serious problem in HRG besides the degree of chemical equilibrium. To date, we assume full equilibrium. Nevertheless, we believe that the HRG model gives an excellent approach, especially for lattice QCD simulations in the hadronic phase. The present analysis leads to limiting temperature, $T_{lim}=164~$MeV. Obviously, this value is slightly higher than the one obtained in Ref.~\cite{Andronic}, $T_{lim}=161\pm4~$MeV. 

The STAR collaboration did not publish results at energies, $27$ and $19~$GeV, yet. Including these measurements would enrich the given data set. STAR BES-II is designed to cover energies $<7.7~$GeV. This would provide another contribution to increase the number of freeze-out parameters and cover larger chemical potential. This would enable us to suggest parametrisation $T_{ch}$ and $\mu_b$ vs $\sqrt{s_{NN}}$ and $T_{ch}$ and $\mu_b$. Also, future facilities NICA and FAIR facilities are supposed to cover region of the QCD phase diagram intersecting with that of BES program. Although, ALICE experiment already published almost the same set of particle ratios. Nevertheless, we did not include this, because of the suppression observed in $p/\pi$ ratio~\cite{ALICE}. Many authors are speculating about the physical reason of this suppression~\cite{Petran,SHMUrQM}.

Confronting the resulting freeze-out parameters with the three conditions  vanishing $\kappa\, \sigma^2$ \cite{Tawfik:2013dba},  $s/T^3=7$  \cite{Tawfik:2005qn,Tawfik:2004ss} and $(\epsilon-3P)/T^4=7/2$ \cite{Tawfik:2013eua}, we conclude that the agreement is not convincing. Nevertheless, vanishing $\kappa\, \sigma^2$ \cite{Tawfik:2013dba} is relatively close, especially at large chemical potentials. To draw final conclusion, we still need further data, at least BES-I and BES-II. 

In a forthcoming work, it intends to study the effect of a certain degree of non-equilibrium, especially the one related for non-equilibrium $\gamma_s$ (at $\gamma_q=1$). Also, we plan to confront recent ALICE results on the particle ratios to the statistical-thermal model. This would a suitable occasion to discuss the suppression observed in $p/\pi$ ratio~\cite{ALICE,Petran,SHMUrQM}.

\section*{Acknowledgement}
This work is supported by the World Laboratory for Cosmology And Particle Physics (WLCAPP), http://wlcapp.net/. The authors are very grateful to Nu Xu, Feng Zhao and Sabita Das for providing them with STAR BES-I particle ratios!

%%%%%%%%%%%%%%%%%%%%%%%%%%%%%%%%%%%%%%%%%%%%%%%%%%%%%%%%%%%%%%%%%%%%%%
%%%   References
%%%%%%%%%%%%%%%%%%%%%%%%%%%%%%%%%%%%%%%%%%%%%%%%%%%%%%%%%%%%%%%%%%%%%%

%-----------------------------------------------

%----------------------==================-------------------------


\begin{thebibliography}{99}
\bibitem{calculated} The experimental ratios at 39, 11.5 and 7.7 GeV, $k^+$/$\pi^+$ and p/$\pi^+$, are calculated using the other experimental ratios assuming the same relative error in $k^-$/$\pi^-$ and $\bar{p}$/$\pi^-$, respectively.
\end{thebibliography}
\end{document}